\documentclass{elsart5p}
\usepackage{graphics}
\usepackage{graphicx}
\usepackage{amssymb}
\def\sgn{\mbox{sgn}}
\begin{document}
\begin{frontmatter}
%
\title{Bose-Fermi Kondo model with Ising anisotropy: cluster-Monte Carlo approach}
\author[AA]{Stefan Kirchner\corauthref{Name1}},
\ead{kirchner@rice.edu}
\author[AA]{Qimiao Si},
\address[AA]{Department of Physics \& Astronomy, Rice University, Houston,
TX 77005-1892, USA}
\corauth[Name1]{Corresponding author. Tel: (713) 348-4291 fax: (713)
348-4150}
\begin{abstract}
The Bose-Fermi Kondo model (BFKM) captures the physics of the 
destruction of Kondo screening, which is of extensive current 
interest to the understanding of quantum critical heavy fermion metals. 
There are presently limited theoretical methods to study the 
finite temperature properties of the BFKM. Here we provide some
of the consistency checks on the cluster-Monte Carlo
method, which we have recently applied to the Ising-anisotropic
BFKM. We show
that the method correctly captures the scaling properties of the Kondo
phase, as well as those on approach to the Kondo-destroying quantum
critical point. We establish that comparable results are obtained
when the Kondo couplings are placed at or away from a Toulouse point.
\end{abstract}
\end{frontmatter}
The sub-Ohmic Bose-Fermi Kondo models are 
of considerable interest in the context of 
quantum critical heavy fermions~\cite{Si.01}
and certain mesoscopic structures~\cite{Kirchner.05}. The finite
temperature scaling properties of the BFKM have been studied
using an $\epsilon$-expansion~\cite{Zhu.02} and also at certain
large-N limit~\cite{Zhu.04}. At finite $N$ and $\epsilon$, standard
Monte-Carlo methods~\cite{Grempel.99} can be used for the Ising-anisotropic
BFKM but the lowest temperature that has been reached is about $0.01~T_K^0$,
where $T_K^0$ is the bare
Kondo scale. Recently, we have applied a cluster-Monte Carlo method
to this problem~\cite{Kirchner.07}, which is able to reliably reach 
temperatures
of the order of $10^{-3}-10^{-4}T_K^0$. The purpose of this paper 
is two-fold. First, we demonstrate the consistency of this method 
for both the Kondo phase and the quantum critical regime.
Second, we address the effect of the deviation
from the Toulouse point on the finite-temperature scaling properties;
previous studies of the Ising-anisotropic BFKM~\cite{Kirchner.07} 
have focused on the
Toulouse point of the Kondo couplings.

The cluster-Monte Carlo method builds on the well-established
understanding~\cite{Yuval.70} that the Kondo problem in the
scaling regime can be studied by a one-dimensional Ising model with 
long-ranged interactions. 
Through a Coulomb-gas picture of spin flips,
the Kondo Hamiltonian, 
$H_{K}=J_{||}S^{z}s_{z}+\frac{1}{2}
J_{\perp}(S^{+}s_{-}+S^{-}s_{+})+H_0(c)$,
is mapped onto a classical Ising chain $H_{I}=\sum_i K_{nn}S^z_i
S^z_{i+1}+\sum_{i<j}K_{lr}^K S^z_iS^z_j$ with algebraically
decaying (long-ranged) interaction 
$K_{lr}^K(r_i-r_j)=K/|r_i-r_j|^{2-\epsilon}$ 
placed at its lower critical dimension
($\epsilon=0$)~\cite{Yuval.70,Grempel.99}. Here, ${\bf S}$ is the
impurity spin and ${\bf s}$ is the electron spin density at the
impurity site of the $c$-electrons with a featureless dispersion
in its kinetic term, $H_0(c)$; $S^z_i$ is an Ising variable at 
the chain site $i$. 
The coupling constant $K_{nn}$ depends on $J_{\perp}$ while $K$ 
is soley a function of $J_{||}$ and can be expressed entirely through
the scattering phase shift $\delta$ of the electrons, $4
\tan{\delta}=\pi J_{||}\rho$, where $\rho$ is the conduction electron
density of states at the Fermi energy.
An extension of this
equivalence has recently been used to address the quantum critical
properties of Ising-anisotropic BFKMs~\cite{Grempel.03,Kirchner.07},
\begin{eqnarray}
{\cal H}_{\mbox{\small bfkm}} =
{\cal H}_K 
+ \; \tilde{g} \sum_{p} S^z \left( \phi_{p} + \phi_{-p}^{\;\dagger}
\right) + \sum_{p} w_{p}\,\phi_{p}^{\;\dagger} {\phi}_{p}\; ,
\label{EQ:H-imp}
\end{eqnarray}
where the impurity spin ${\bf S}$ interacts with fermions
$c_{p\sigma}^{\dagger}$ and bosons $\phi_{p}^{\;\dagger}$.
The spectrum of the bosonic bath is taken to be
sub-Ohmic ($0<\epsilon<1$),
$
\sum_p [\delta(\omega-\omega_p)- \delta(\omega+\omega_p)] \sim
|\omega|^{2-\epsilon} \sgn(\omega)$
and gives rise to an interaction $K_{lr}^g(r_i-r_j)=g/|r_i-r_j|^{2-\epsilon}$
along the Ising chain on top of the
$K_{lr}^{K}(r_i-r_j)=K/|r_i-r_j|^{2}$ for the Fermi-only Kondo model.

Before discussing the determination of the critical coupling
$\tilde{g}_{c}$ or $g_c$, it is worthwhile discussing the $g=0$ case in
greater detail. The RG flow of the Kondo model is towards an SU(2)
invariant fixed point on trajectories with
$J_{||}^2(b)-J_{\perp}^2(b)=c$, where $b$ parametrizes the RG flow and $c$
is a positive constant. At the Toulouse point, the Kondo model can
be mapped onto a resonant level model~\cite{Guinea.85}. 
Deviation from the Toulouse point is needed to restore the 
SU(2) symmetry at the fixed point.
The Ising chain $H_I$ at $\epsilon=0$ does not possess a
continuous symmetry but undergoes a Kosterlitz-Thouless-like 
phase transition; the RG flow resembles that of the Kondo
model away from the Toulouse point. 
This raises the question of whether the universal properties
of the BFKM, obtained from simulating the
classical Ising chain,
are sensitive to whether
the Kondo model is placed at or away from its Toulouse point.
Previous simulations have utilized the Toulouse point, 
$(1-2\delta/\pi)^2=1/2$, where a comparison with exact results
is possible~\cite{Grempel.99,Kirchner.07}. 
Here we present 
results away from the Toulouse point of  $ {\cal
  H}_{\mbox{\small bfkm}}$ in Equation (\ref{EQ:H-imp}) with
$\tilde{g}=0$. We choose $(1-2\delta/\pi)^2=3/5$, but keep
$J_{\perp}=0.75\tau_0$ at the value considered in~\cite{Kirchner.07} for easy
comparison. This results
in a bare Kondo temperature $T_K^0=0.553\tau_0$, with a Trotter parameter
$\tau_0=\beta/L=1/16$. 
\begin{figure}[t!]
\begin{center}
\includegraphics[angle=0,width=0.415\textwidth]{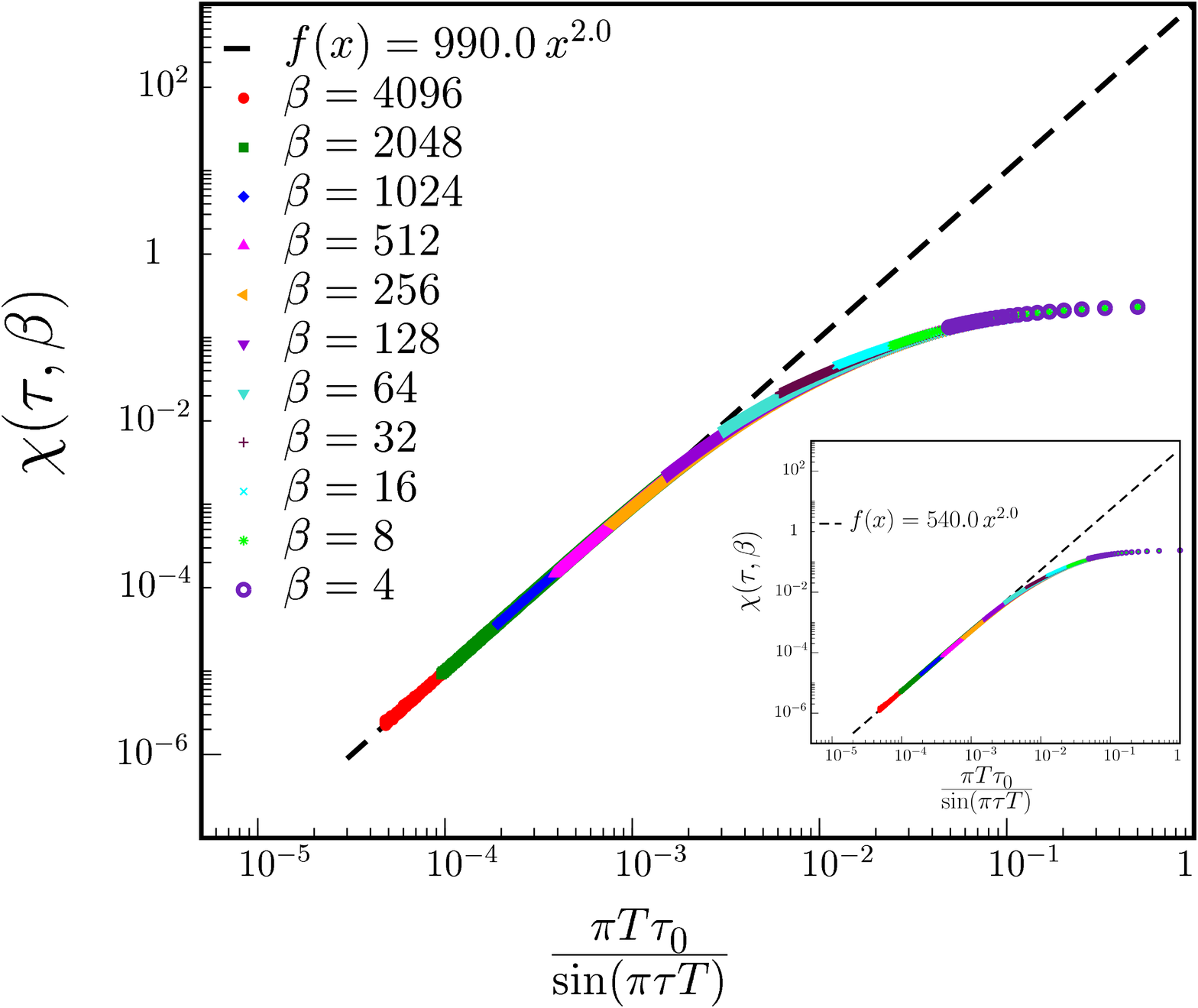}
\end{center}
\caption{The scaling function for the Kondo model away from its
  Toulouse point: $J_{\perp}=0.75\tau_0,(1-2\delta/\pi)^2=3/5$ and $\tau_0=1/16$.
The inset shows our results right at the Toulouse point, 
$(1-2\delta/\pi)^2=1/2$; all other couplings, and the temperatures,
are the same as in the main plot.
} \label{fig1}
\end{figure}
The Ising chain  was solved via a Cluster-MC
scheme~\cite{Kirchner.07,Luijten.95}. Fig.~\ref{fig1} displays the results 
for the dynamical susceptibility $\chi(\tau,T)=<T_{\tau}S^z(\tau) S^z(0)>$
for various temperatures. 
All
curves collapse on one scaling curve $W(x)$ with $W(x) \sim x^2$ for
$x\ll T_K$ when plotted as a function of the
single variable $x=T/\sin({\pi \tau T})$. This  is a consequence of the
conformal invariance of the Kondo Hamiltonian [Equation (\ref{EQ:H-imp}) with
$\tilde{g}=0$]. This result immediately implies the correct frequency
and temperature dependences of the dynamical susceptibility. 
A similar scaling collapse applies
at the Toulouse point, as seen in the inset.

\begin{figure}[t!]
\begin{center}
\includegraphics[angle=0,width=0.415\textwidth]{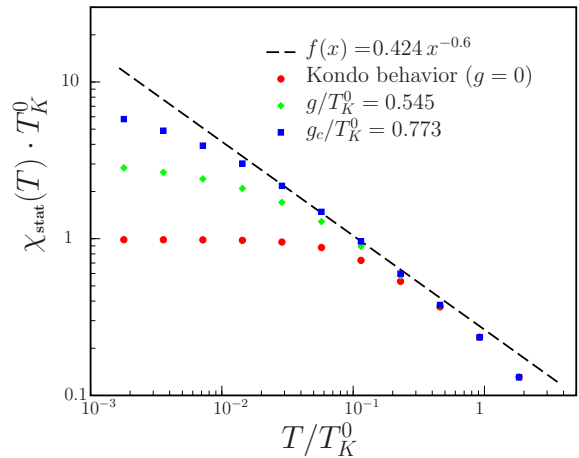}
\end{center}
\caption{Static susceptibility for 
$\tau_0=1/16$, $J_{\perp}=0.75\tau_0$, $(1-2\delta/\pi)^2=3/5$ versus
temperature for the Kondo model $g=0$, at an intermediate
coupling $g/T_K=0.545$ and at the critical coupling
$g_c/T_K=0.773$ for $\epsilon=0.4$. The divergence of $\chi(g_c)$ is cut
off due to the finite system size $L=\beta \tau_0^{-1}$;
the cutoff temperature decreases as $\tau_0$ decreases~\cite{Kirchner.07}.
} \label{fig2}
\end{figure}

The static susceptibility versus
temperature $T$ for the Kondo model [$g=0$,
away from Toulouse point with
$(1-2\delta/\pi)^2=3/5$]
is shown in Fig.~\ref{fig2}. These results in conjunction with the
results from reference~\cite{Kirchner.07} establish that the universal
properties obtained via our method are insensitive to whether
the Kondo model is placed at or away from its Toulouse point.
We can now study the approach to the Kondo-destroying QCP at $g=g_c$.
As $g$ increases, the static local susceptibility deviates from 
the Pauli behavior of the
Kondo phase. As $g$ approaches $g_c$, an intermediate temperature
range opens up, which shows a power-law divergence. The critical
exponent is consistent with the perturbative RG result,
$\chi(T) \sim T^{-(1-\epsilon)}$. The lower cutoff temperature 
of this critical regime becomes smaller as $g$ increases towards
$g_c$, consistent with the general expectations of 
quantum critical scaling.
$\chi(\tau,T)$ shows a scaling
collapse similar to that shown in Fig.~\ref{fig1}, 
providing evidence for an enhanced
symmetry at $g=g_c$, the quantum critical point~\cite{Kirchner.07}.

In summary, we have established that the cluster-Monte Carlo
study of the classical Ising chain with long-range interaction
can be used to  obtain universal properties of the Ising-anisotropic
Bose-Fermi Kondo model,
both at and away from the Toulouse point.
Because of the large temperature range it is able to reliably
access, the approach is expected to be useful for studying
the quantum critical behavior of not only the BFKM itself,
but also the extended dynamical mean field theory of the 
Kondo lattice model.

This work has been supported in part by
NSF, the Robert A. Welch Foundation, the W. M. Keck Foundation,
and the Rice Computational Research Cluster
funded by NSF
and a partnership between Rice University, AMD and Cray.


\begin{thebibliography}{1}
\vspace*{-0.25cm}
\expandafter\ifx\csname url\endcsname\relax
  \def\url#1{\texttt{#1}}\fi
\expandafter\ifx\csname urlprefix\endcsname\relax\def\urlprefix{URL }\fi
\bibitem{Si.01}
Q.~Si et al., Nature 413 (2001) 804.
\bibitem{Kirchner.05}
S.~Kirchner et al., PNAS 102 (2005) 18824.
\bibitem{Zhu.02}
L.~Zhu and Q.~Si, Phys.~Rev.~B 66 (2002) 024426;
G.~Zarand and E.~Demler, {\it ibid.} (2002) 024427.
\bibitem{Zhu.04}
L.~Zhu et al., Phys.~Rev.~Lett. 93 (2004) 267201.
\bibitem{Grempel.99}
D.~Grempel and M.~Rozenberg, Phys.~Rev.~B 60 (1999) 4702.
\bibitem{Kirchner.07}
S.~Kirchner and Q.~Si,
Phys.\ Rev.\ Lett. in press and arXiv:0706.1783v1.
\bibitem{Yuval.70}
G.~Yuval and P.~W. Anderson, Phys.~Rev.~B 1 (1970) 1522.
\bibitem{Grempel.03}
D.~Grempel and Q.~Si, Phys.~Rev.~Lett. 91 (2003) 026401.
\bibitem{Luijten.95}
E.~Luijten and H.~W.~J. Bl\"ote, Int.~J.~Mod.~Phys. 6 (1995) 359.
\bibitem{Guinea.85}
F.~Guinea et al., Phys.~Rev.~B 32 (1985) 4410.
\end{thebibliography}
\end{document}